\documentstyle[twocolumn, epsfig,11pt]{article}
\makeatletter \makeatletter
\def\@normalsize{\@setsize\normalsize{10pt}\xpt\@xpt
\abovedisplayskip 10pt plus 2pt minus 5pt
\belowdisplayskip
\abovedisplayskip\abovedisplayshortskip \z@ plus 3pt
\belowdisplayskip 6pt plus 3pt minus 3pt\let\@listi\@listI}
\def\subsize{\@setsize\subsize{12pt}\xipt\@xipt}
\def\section{\@startsection{section}{1}{\z@}{1.0ex plus 1ex minus 0.2ex}
{0.2ex plus 0.2ex}{\large \bf}}
\def\subsection{\@startsection{subsection}{2}{\z@}{0.2ex plus 1ex}
{0.2ex plus 0.2ex}{\subsize \bf}}
\makeatother

\small

\begin{document}

\baselineskip 11pt
\title{\Large \bf Chaos for Stream Cipher}

\author{\normalsize
    \begin{tabular} {cc}\\
    {Ninan  Sajeeth Philip} & {K. Babu Joseph}\\\\ 
    {Department of Physics} & {Department of Physics}\\ 
    {Cochin University ~of} & {Cochin University ~of}\\ 
    {Science and Technology}& {Science and Technology}\\ 
    {Kochi - 682 022, India}& {Kochi - 682 022, India}\\ 
    {\em nsp@stthom.ernet.in}& {\em vc@cusat.ac.in}\\
    \end{tabular} }
    \date{}

\maketitle
\subsection*{\centerline{\large\bf Abstract}}

{\em This paper discusses mixing of chaotic systems as a dependable
method for secure communication. Distribution of the entropy function
for steady state as well as plaintext input sequences are analyzed. It is
shown that the mixing of chaotic sequences results in a sequence that
does not have any state dependence on the information encrypted by
them. The generated output states of such a cipher approach the
theoretical maximum for both complexity measures and cycle length.
These features are then compared with some popular ciphers.}\\

{\bf{Keywords:}} chaos, stream ciphers, logistic equation, mixing of chaos \\

\section{Introduction}

With the advent of the world wide web and e-commerce, research in
digital cryptography \cite{Bras88} has acquired a renewed momentum.
In general, the algorithms used for cryptographic applications may be
classified into two, namely, public key cryptography and secret key
cryptography. Public key cryptography uses separate keys to encrypt
and decrypt an information. In this system a person uses his private
key and the public key of the legitimate recipient to encrypt a message.
At the receiving side, the message is decoded using the public key of
the sender and the private key of the recipient. The advantage of the
scheme is that anyone can send an information securely to the person
who has a public key and make sure that only the addressee will be able
to read it. Thus they are mainly used for secure exchange of information
between two persons who need not trust each other. Public key
cryptosystems accomplish this by relying on the inability of present day
tools to factor large numbers (eg. RSA like implementations) or on the
difficulty in solving the discrete logarithm problem (Diffie-Hellman like
implementations) etc. \cite{Diffie79}. Since such computations take
reasonable time on most implementations, they become vulnerable to a
host of attacks. The nature of the key may be understood by monitoring
the time taken for encryption and decryption. This is the basis of \textit
{timing attack}. A slightly different attack known as the \textit
{differential power analysis} monitors the power consumed by the
encoding devices and deduce information on the secret key. A more
generalized attack known as \textit {differential cryptanalysis} maps the
correspondence between the plain text and cipher text symbols (state
vectors) to decode the secret key. On the other hand, secret key
cryptography, popularly known as stream ciphers use the same key for
encryption and decryption. Since this demands the secure exchange of
secret keys between the users, their role is slightly different from that
of public key cryptosystems. They are mainly used for the secure
storage and retrieval of information. For example, in the banking sector
or in the credit card application, there is no need for a public key. Since
the user need not have to declare any public key, these ciphers are
generally more secure compared to public key cryptosystems. Most of
the stream ciphers use simple logical operations that are very fast
compared to public key cryptosystems and are thus immune to
differential power analysis or timing attacks. However, they could be
vulnerable to \textit{differential cryptanalysis}, \textit {brute force
attacks} and \textit {known plaintext attacks} unless necessary
precautions are taken.

Most stream ciphers rely on random number generators for encryption.
At the transmitter side the secret information is added to a random
carrier signal. At the receiving end it is subtracted out from the received
signal to retrieve the information. Since a random process cannot be
reproduced with certainty, it is not possible to generate the same
random sequence at both the transmitter and receiver simultaneously.
The general practice is to use pseudo-random numbers that can be
generated at both sides without difficulty. 

Studies in nonlinear dynamics shows that many of the seemingly
complex systems in nature are described by very simple mathematical
equations \cite{Berge, Gleick, Nicolis}. Fractal images, for example, are
capable of traversing the embedding space in minimum time, a feature
that is referred to as the space filling ability of fractal images. Viewed
from the cryptographic point of view, this is equivalent to the maximum
cycle length \cite{RitterT}, a feature expected from a good random
number generator (RNG). Although chaotic systems appear to be highly
irregular, they are also deterministic in the sense that it is possible to
reproduce them with certainty. These promising features of chaotic
systems attracted many researchers to try chaos as a possible medium
for secure communication. Chaos was successfully implemented in
analog secure systems following the works of Pecora and Carroll in
1990 \cite{Pecora, Hong, Young}. However, implementations in digital
systems have not yet been successful. The major reason why digital
cryptography using chaos did not turn out to be effective is that digital
circuits deal with finite number spaces. Chaos is a phenomenon in which
even very small variations in the initial state of a system results in
exponentially diverging evolution of its future states. When dealing with
finite number space, chaos looses this diversity due to the restricted
resolution of possible states. However, many researchers have pointed
out that mixing of pseudo random sequences of finite precision are able
to produce rather complex sequences
\cite{RitterT90,MacLaren,Wolfram,Pless}. Their studies were however
largely limited to linear systems. The present paper is an extension of
these encouraging observations to nonlinear chaotic systems that have many
additional features suitable for cryptography.

The organization of the paper is as follows: In section \ref{mcs} we
introduce the new cipher that generates a space filling sequence by
mixing two chaotic sequences. Section \ref{software} discusses the
software implementation of the cipher. Section \ref{cryptan} presents
some popular cryptanalysis techniques on the proposed system. Some
methods to investigate the security aspects of a cipher in general are
also discussed. Finally, Section \ref{popular} compares the new cipher
with three popular ciphers. \\

\section{Mixing of chaotic sequences\label{mcs}}

One major difference between digital and analog systems is the introduction
of round-off errors in digital systems. This is observed to give some peculiar
effects that are explained in detail in \cite{Beck}. On the other hand, the
assurance of exact reproducibility is an added advantage of digital systems.
Studies on digital chaos generators have been reported to have more diversity
compared to their analog counterparts \cite{Cernak}. This is encouraging, since
cryptographic applications are mostly on digital platforms. 

Matthews \cite{Matthews} in 1989 proposed an encryption algorithm
using the principles of chaos. He generalized the logistic equation (see
equation \ref{logistic}) with cryptographic considerations and used the
last two decimal digits of the generated sequence to get the random
sequence. This system was found to have various drawbacks as
pointed out by Wheeler \cite{Wheeler}. Matthews' algorithm gave way to
one-way functions \cite{Mitchell} in 1990. 

A common route to chaos in most systems is through period doubling
\cite{Berge}. In the fully developed chaotic system, there exist
simultaneously a large number of unstable periodic trajectories for the
evolution of the system. At this stage even minor fluctuations in their
initial conditions would cause the system to switch between different
trajectories. Although the digital systems are restricted by the finite
discrete space, it may be shown that a proper shuffling of these states
can generate pseudo random sequences with many interesting
properties. This is in agreement with the observations made by other
researchers as it would be described below.

For the discussion of the proposed method of shuffling or mixing of chaotic
states of a system, we take the logistic equation as an example. The logistic
equation is an iterative equation in which every upcoming state \( x_{n+1} \)
is defined as a nonlinear function of the present state \( x_{n} \) as: 

\begin{equation}\label{logistic}
x_{n+1}=4\lambda x_{n}(1-x_{n})\quad with\quad 0\leq \lambda \leq 1
\end{equation}

\begin{figure}
\centering {\resizebox*{0.5\textwidth}{0.3\textheight}
{\includegraphics{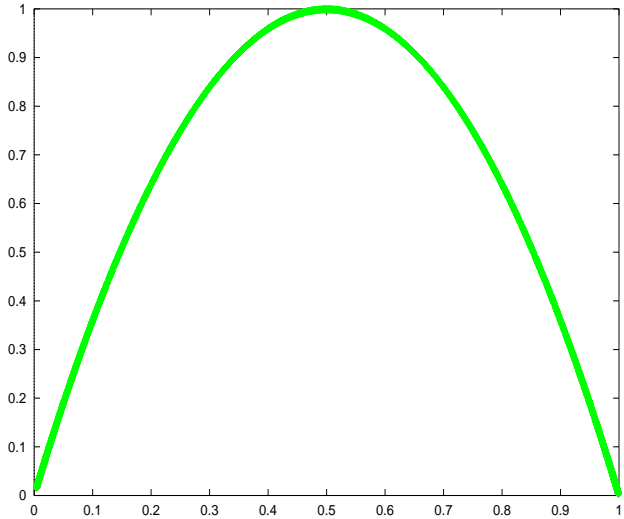}}} \caption{The attractor of the logistic
equation is an inverted parabola. The peak of the parabola corresponds
to the value of the control parameter \( \lambda  \)}\label{fig:fig1a} 
\end{figure}

If a graph is plotted with different values for the control parameter, \(
\lambda  \), many interesting properties of non-linear systems may be
observed. When the value of \( \lambda  \) is close to zero, the sequence
converges to some steady value after a few iterations. On increasing
the value of \( \lambda  \) it enters a periodic cycle. Increasing the value
of \( \lambda  \) further results in a feature known as period doubling
and finally, at values of \( \lambda  \) close to one, the sequence
appears to be completely random. At this stage, even slight variations in
the value of \( x_{n} \) at any instant would result in very large
(exponential) variations in the subsequent evolution of the system. This
is referred to as chaos \cite{Berge}. It should be noted that in the
logistic equation, the \( x_{n+1} \)th state is completely defined by x\(
_{n} \). This is what is known as dependence on initial conditions.
Although the sequence of values generated by the system at this stage
appears to be random, they are in fact highly deterministic as can be
seen from their attractor structure shown in figure \ref{fig:fig1a}. This
makes the system insecure for cryptographic applications.

To study the effect of mixing, let us consider two separate logistic
equations. Let \( x_{n} \) and \( x_{n}^{\prime } \) be two points on the
nearby trajectories of such two systems. For simplicity we assume that
the control parameter \( \lambda  \) is the same for both the systems.
We now define the evolution of a new system by the following sequence of
equations:

\begin{equation}
\begin{array}{c}
x_{n+1}=4\lambda x_{n}(1-x_{n})\\
x^{\prime }_{n+1}=4\lambda x^{\prime }_{n}(1-x^{\prime }_{n})\\
P_{n}=x_{n}\oplus x^{\prime }_{n}\\
x_{n}=x_{n+1}\oplus C_{n}\\
x^{\prime }_{n}=x^{\prime }_{n+1}
\end{array}
\end{equation}
 \label{eqn2}

The \( \oplus  \) operator represent the XOR operation. We refer to
equation (\ref{eqn2}) as the cipher equation in the rest of the paper. 

\begin{figure}
\centering {\resizebox*{0.5\textwidth}{0.3\textheight}
{\includegraphics{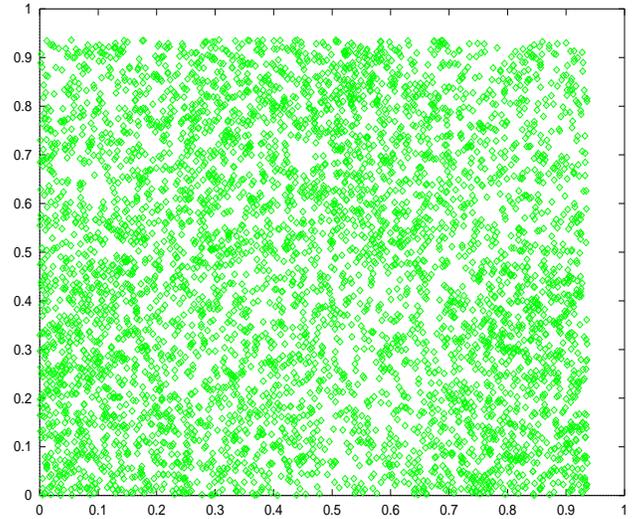}}}\caption{A fully developed chaotic system
has many unstable states of excitation. The mixing of two nearby
trajectories of the logistic equation destroy the attractor structure and
the correlation of adjacent states. This is identical to a continuous
perturbation on an unstable system.}\label{fig:fig1b}
\end{figure}

As mentioned earlier, combining adjacent states of a system using XOR
function is not new in cryptography. Ritter for example present the
concept of pseudo-random shuffling \cite{RitterT90} as a very efficient
method for random sequence generation. MacLaren \emph{et al.}
\cite{MacLaren} showed that combining two congruential generators
would give a more complex random sequence than individual
congruential generators. In cellular automata, Wolfram \cite{Wolfram}
suggested a linear array of elements a{[}0{]}...a{[}n{]} and an iterative
equation a{[}i{]} = (a{[}i-1{]}) XOR (a{[}i{]} OR a{[}i+1{]}) to construct a
RNG. The output was taken as a portion of the state vector. An
analysis of such systems was made by Rietman \cite{Rietman}. Many
combiners using linear feedback shift registers (LFSR) have also been
attempted as alternate sources of random number generators
\cite{Pless}. However, these were on linear systems and due to their
inherent linearity, they become vulnerable to a known plain text attack
or correlation attack \cite{Siegenthaler}. In the proposed system, the
mixing is done on nonlinear systems at the onset of chaos. Thus the
mixing is equivalent to a continuous perturbation on a chaotic system
causing the system to jump from one unstable trajectory to another in
accordance with the perturbation. The system is still deterministic as
long as the nonlinear equations are defined. However, the sequence of
iterative equations produces a high order Markov process due to the
shuffling process in the finite space \cite{Knuth81} making the
generated sequence appear as random to the intruder. This is illustrated
in figure \ref{fig:fig1b} where instead of the attractor, we have a space
filling \textit{noise like} pattern in the phase space. The points in the plot
will be identical in two instances if and only if the encrypted text and
the initial conditions of the logistic equations are all the same. It is
possible to prevent the occurrence of such a situation by initializing the
logistic equation by some non-repetitive parameter, e.g. a psedo
random number generated using the time of the day as the seed.
Although such a parameter appears to be insecure, this is not a serious
issue since the actual security of the system is not based on the value
of this parameter. Even a slight deviation in the value of this parameter
causes the evolution of the system to diverge away exponentially from
each other, an inherent property of chaotic systems, making an entirely
new sequence even when the plaintext contents are the same.

Also note that in the iterative formula suggested in the cipher equation,
\( C_{n} \) is the pseudo random cipher text and is defined outside
equation \ref{eqn2}. For the purpose of this paper, we will define \(
C_{n}=P_{n}\oplus y_{n} \) with \( y_{n} \) as the plaintext data stream.
We will describe a multiple encryption scheme later in the paper which
can improve the security of the system further by an exponential factor.\\

\section{Implementation\label{software}}

In the software implementation, the starting value x\( _{1} \) was
selected as a pseudo random variable and x\( _{1}^{^{\prime }} \) was
computed from x\( _{1} \). Since the value of x\( _{1} \) is required to
reconstruct the sequence, it was transmitted along with the cipher text.
A dummy sequence of random numbers, the length of which is agreed
by the legitimate users, were used to hide the location of x\( _{1} \). It
is possible to adopt any steganographic methods to hide the starting
value, for example, as the \( n \)th bit of a set of bits in a dummy
sequence. The data stream was then grouped into blocks of cipher texts
with a set of synchronizing pulses to ensure proper alignment. All these
signals are encrypted that they appear exactly like the encrypted
information signal to the intruder. Grouping into blocks and
synchronization pulses are not required by the algorithm, but are
included to show that such locations may be used to place tracking
codes, error correction codes or CRC codes. \\

\section{Cryptanalysis\label{cryptan}}

From cryptographic considerations, three measures on the ciphertext
contribute significantly towards the security of the information ciphered
by them. They are the Combiner Correlation \cite{Ritter_COMBCORR},
the Cycle-Length \cite{RitterT} and the Complexity
Measure \cite{Fischer}. Combiner Correlation estimate the probability
with which the output of the cipher becomes identical to that of the
input \cite{Ritter_COMBCORR}. In our example, it is the correlation
between
\( y_{n} \) and \( C_{n} \) . If for example the output follows the input by
an unequal number of times, this information may be used to break the
cipher. The simplest way to compute this correlation is by writing a
table of the possible inputs and the corresponding outputs. For
simplicity, it is assumed that the variables are of single bit boolean
type. In table \ref{table:tab1}, for single variables like
\( x_{n} \), their actual value is indicated, while for comparison of two
variables, we put a \( 1 \) to indicate that the variables are equal. For
example \( y_{n}=C_{n} \) has a value \( 1 \) to indicate that they are
equal.  As it can be seen (\textit{last column}), the cipher text has equal
probability of occurrence when the values of \(x_n, x^{\prime}_n\) and
\(y_n\) flips their states. Since the cipher equation uses the XOR
operation, which is a bit level mod 2 addition operation, the same
pattern repeats even when the variables are multiple bit words. Thus
the proposed system will not leak out any information regarding the
possible input state vectors in the cipher text.\\
\begin{table}[h]
\centering
\begin{tabular}{|c|c|c|c|c|c|}
\hline 
\( x_{n} \)&
\( x^{\prime }_{n} \)&
\( P_{n} \)&
\( y_{n} \)&
\( C_{n} \)&
\( y_{n}=C_{n} \)\\
\hline 
0&0&0&0&0&1\\
\hline 
0&0&0&1&1&1\\
\hline 
0&1&1&0&1&0\\
\hline 
0&1&1&1&0&0\\
\hline 
1&0&1&0&1&0\\
\hline 
1&0&1&1&0&0\\
\hline 
1&1&0&0&0&1\\
\hline 
1&1&0&1&1&1\\
\hline 
50\%&50\%&50\%&50\%&50\%&50\% \\
\hline 
\end{tabular}\caption{The table shows the possible distribution of the
cipher text and the intermediate states for the possible values of \(
x_{n} \) and \( x^{\prime }_{n} \).}\label{table:tab1}
\end{table}

\subsection{Differential Cryptanalysis}

Differential cryptanalysis \cite{Biham90} makes use of some chosen
plain texts to explore any possible correlation between the plain text and
the cipher text state vectors. It is difficult to prove that a cipher would
be equally strong in hiding vital information about itself irrespective of
the plain text patterns enciphered by it. However, there are a few
standard tests to verify such possibilities. One test is to look for the
existence of indicative variations in the probability distribution density
of the output states of the cipher text as a function of the plain text
information masked by it. A reliable measure of this is the cycle length
\cite{RitterT} of the ciphertext when the plain text consists of only one
kind of symbol. Popular measures on cycle length such as the run-up
tests are difficult on self-synchronous ciphers of the type discussed
here. We thus do an indirect measurement of this feature. Information
theoretical computations such as entropy measures are based on the
probability distribution of state vectors in a sequence. A uniform
distribution of state vectors at all possible values in a finite space is
equivalent to a maximum cycle length since such a distribution demands
the system to traverse all possible states before repetition. This can be
directly measured as the probability of occurrence of each state vector
in the finite space when the length of the sequence is in the order of the
total number of possible states. 

Two different sets of ASCII character sequences were used for the analysis.
In the first round of tests, the sequences consisted of each with a single ASCII
character repeated 9464 times. We say that the information is masked by the
cipher, if it is not possible to identify any specific trend in the cipher corresponding
to the distribution in the plain text. For ease of comparison, we computed the
entropy distribution of the sequences. If the probability distribution density
of the plain text character \( x \) in the cipher text is distributed over
a range of M possible states and if the population density in each state is
\( P_{j} \), then the Shannon entropy of the distribution in unit bits is given
by

\[
H_{x}=-\sum ^{j=M}_{j=1}P_{j}*log_{2}P_{j}\]

\begin{figure}[h]
\centering {\resizebox*{0.5\textwidth}{0.3\textheight}
{\includegraphics{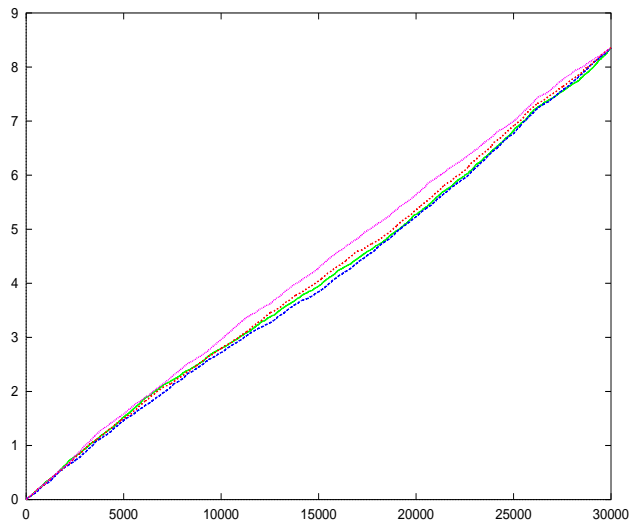}}}\caption{The entropy distribution of the
proposed system as a function of bit count. The entire 16 bit number
space was divided into 30,000 equal slots. The distribution is for eight
bit ASCII characters represented by decimal digits 01, 30, 31 and 255.
Note the linear increase in the entropy distribution as a function of the
bit count. This indicates a uniform distribution at all bit levels and hence
a maximum cycle length for the random sequence.}\label{fig:fig2a}
\end{figure}
\begin{figure}[h]
\centering {\resizebox*{0.5\textwidth}{0.3\textheight}
{\includegraphics{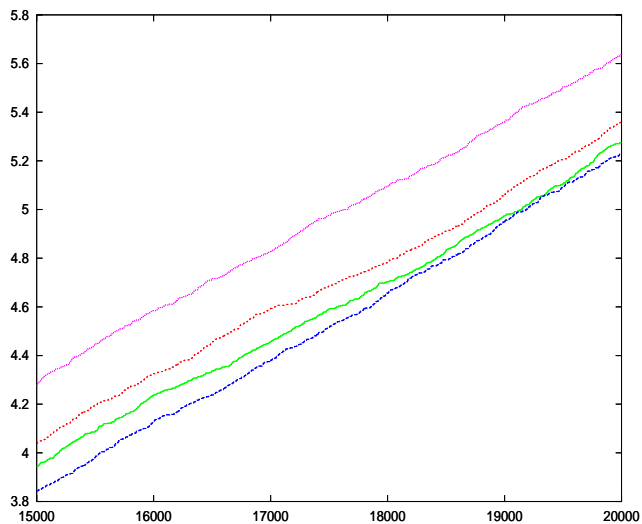}}}\caption{A magnified view of the central
region of figure \ref{fig:fig2a} shows a marginal deviation in the entropy
distribution for each ASCII value. Although it appears to reveal some
information regarding the plain text information, it remains to be known
how far this could be used as a possible tool for breaking the cipher
equation.}\label{fig:fig2b}
\end{figure}

The graphical representation of this measure reveals any orientation of
the distribution density towards the corresponding plain text
information. The distribution of the entropy of the cipher texts
corresponding to the ASCII characters represented by the decimal digits
01, 30, 31 and 256 are shown in figure \ref{fig:fig2a}. An enlarged portion
of ~figure \ref{fig:fig2a} is shown in figure \ref{fig:fig2b}. Figure
\ref{fig:fig2a} indicates that the entropy converges to the same value
while its deviates slightly in the middle. Also it is observed that the
entropy distribution increases linearly as a function of the~bit-count.
This means there is a uniform distribution of the random values over the
entire state vector space. In other words, the cipher has effectively
masked the information by generating a space filling sequence with
uniform probability of occurrence at all levels, which is ideal from a
theoretical point of view. 

In the second round of tests, we used random sequences of English
prose of equal length as plain text. It was observed that the entropy
curves were more linear and indistinguishable compared to that of the
single variable sequences. 

In both cases, the linearity of the curve ensures maximum cycle length
and uniform distribution over all the possible state vectors \cite{Fischer},
while the non-linearity of the chaotic system introduces necessary
randomness. Thus we expect the cipher to be reasonably secure against
differential cryptanalysis.\\

\subsection{Complexity measure}

The widely accepted test for the randomness of a sequence is the
Chi-square test \cite{Knuth81}. The cipher described in the present
paper uses the cipher text itself as part of its input vectors. As a
consequence, just like other self-synchronous stream ciphers, it is
difficult to judge this cipher using Chi-square tests. 

However, the randomness of a sequence may be estimated from its
complexity. By complexity we mean the minimum number of bits
required to represent the information contained in it. A direct measure of
this is its compressibility \cite{RitterT}. Although the sequence is
pseudo random, meaning that they may be compressed using the exact
equation that generated them, since the details of the parameters of the
equation (secret keys) are not available to the hacker, we discard this
possibility. In this situation, a rough estimate of the complexity of a
data set may be obtained using the Huffman algorithm \cite{Huffman}
for data compression. The Huffman algorithm computes the redundancy
of data in a sequence and then replaces them using some symbols that
require lesser number of bits and thus reduces the size of the data
sequence. We do not go into the details of the algorithm here, but would
cite \cite{Heller} as a practical reference. The ratio of the size of the
original cipher text to that of the compressed text is a measure of the
complexity or randomness of the cipher text. The higher the ratio, the
lesser is the complexity. 

For the analysis, a large set of English prose was chosen at random.
Each plain text was about a few kilobytes in size. The choice was made
taking into consideration the ease of comparison with earlier studies. We
used the ASCII code to represent the English alphabets (8 bits per
letter). The Linux version of \textit{gzip software} with maximum
compression option (gzip -9) was used to compress the files. This is
the GNU lossless data compression package popular on most UNIX type
machines. In all the test cases, the compression ratio was less than one
since the coding resulted in files larger in size than the original files. This
additional size was required to keep the coding information and the
original file could not be compressed at all. This indicates that the
complexity measure of the cipher text sequence is the same as the
complexity of the embedding finite space and is the theoretical
maximum. \\

\subsection{Brute force attack}

The straight forward attack on any cipher is the brute force attack.
Except for the computational intensity, this is as simple as trying all the
possible combinations of the variables in the system. In the cipher
proposed by the cipher equation, the independent variables are \(
\lambda  \), \( x_{n} \) and \( x^{\prime }_{n} \). Assume that the value
of \( \lambda  \) is significant up to k bits and that of \( x_{n} \) and \(
x^{\prime }_{n} \) each up to m bits. We say that a bit is significant if
the system has a different evolution when the bit flips its value. Thus
for brute force attack, the hacker requires a maximum of \( 2^{2m+k} \)
attempts to break the value of \( \lambda  \). If we assume that \(
k=m=16 \) bits, the brute force attack requires a maximum of \( 2^{48}
\) computations. 

The general procedure of the attack is like this: take some samples from
the middle of the ciphertext that appears to have the highest probability
of containing some meaningful data. Reverse engineer using the cipher
equation all the possible values of \( x_{n} \), \( x^{\prime }_{n} \) and \(
\lambda  \) until some meaningful text is generated. If the hacker has a
known plain text database, the complete process may be automated.
Then he only need to compare the two. If an exact match is found, he
has succeeded in breaking the codes. He may now apply this
information on the entire ciphertext and retrieve the location and method
used to hide the starting value. This completely decodes the cipher. 

One solution to this problem is to increase the number of bits used for
the parameters in the cipher. But this is not foolproof, since the
probability for getting the right guess in a reasonable number of
iterations is not zero. We suggest multiple encryption as a better
approach. Since a single flip of any of the bits of the parameters would
result in a different sequence that diverge away from the original
sequence exponentially at every point, in this case, the hacker has to
either assume the ciphertext generated in the previous encryption or
has to do \( 2^{n(2m+k)} \) computations to obtain the plaintext
information. Here \( n \) is the number of times the encryption has been
done. Certainly, there is an upper limit to the possible maximum value
for \( n \). This is mainly due to the finiteness of the embedding phase
space and the block size. As computational power increases, there
would be a proportional increase in the word size handled by the
machine also. Thus it may be argued that there would always exist a
safe margin between the attainable complexity and the time required for
brute force attacks. \\

\subsection{Plain text attack}

We will now attempt a known plain text attack on the system. Let us
assume that P\( _{n} \) represent a k-bit number with each bit
represented by P\( _{n} \){[}k{]}. Then, P\( _{n} \){[}k{]} = 0 iff x\( _{n}
\){[}k{]} = x\( _{n}^{^{\prime }} \){[}k{]}. The initial setting is such that \(
x_{0} \) and \( x^{\prime }_{0} \) are different. However, due to the
mixing of \( x_{n+1} \) with \( C_{n} \), there is a non-zero probability
for \( P_{n}=0 \). The risk with \( P_{n} \) going to zero is that this would
cause the cipher to be same as the plain text information. However, it
can be shown that \( P_{n} \) and \( P_{n+1} \) will never be zero
simultaneously if \( y_{n} \), the plain text is not a null vector. This is a
simple condition since a null vector does not have any information
content and is never transmitted.

In this situation, with a known plain text, the attacker is able to identify
the locations in the cipher text that match the state vectors in the plain
text and thus conclude that they correspond to locations where \(
P_{n}=0 \). Considering this possibility, the question is to what extent
the attacker may extend this information to crack the secret codes,
namely \( \lambda  \) and some other critical parameters used to
generate the Cipher text like the coding of the starting vector \( x_{0} \). 

Let at some level \( m \), \( P_{m}=0 \). 

This means, \( x_{m}=x^{\prime }_{m} \). 

\( \Rightarrow  \)\( x_{m}^{*} \bigoplus x_{m-1} \bigoplus x_{m-1}^{^{\prime }} \bigoplus y_{n}=x^{^{\prime }}_{m} \)
~with \( x^{*}_{m} \) representing \( x_{n+1} \) in the cipher equation. 

\( \Rightarrow  \)\( x_{m}^{*}\bigoplus x_{m-1} \bigoplus x_{m-1}^{^{\prime }} \bigoplus x^{^{\prime }}_{m}=y_{n} \)

\( \Rightarrow  \)\( (4\lambda x_{m-1}(1-x_{m-1})) \bigoplus x_{m-1} \bigoplus x_{m-1}^{^{\prime }} \)
\( \bigoplus (4\lambda x^{^{\prime }}_{m-1}(1-x^{^{\prime }}_{m-1}))=y_{n} \)\\

Due to the inability to solve the above equation for \( \lambda  \) and
\( x_{0} \), we conclude that the cipher is resistant to known plain text
attacks. \\

\section{Comparison with other Ciphers\label{popular}}

For a comparison of the new method with some of the popular ciphers, three popular
ciphers were chosen. They are: 
\begin{enumerate}
\item Crypto (v.2.9) 95|NT: Copyright 1998 by Gregory Braun. This software is a
free-ware. 
\item PGP 5.0 for Personal Privacy, Evaluation version, Copyright 1997 by Pretty
Good Privacy and using RSA key of size 2048. 
\item PGP 5.0 for Personal Privacy, Evaluation version, Copyright 1997 by Pretty
Good Privacy and using DSS/Diffie-Hellman key of size 1024/3072. 
\end{enumerate}

\begin{figure}
\centering {\resizebox*{0.5\textwidth}{0.3\textheight}
{\includegraphics{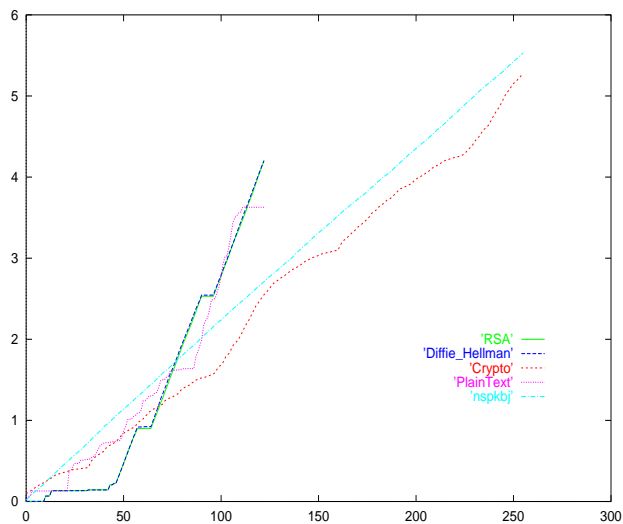}}}\caption{A comparative study of the proposed system with three popular ciphers
is shown here. A flat region in the graph represents regions that does not have
representative vectors in the cipher text. The ideal case is to span the entire
phase space uniformly ( a straight line with positive slope), as that would
correspond to maximum cycle length for the generated random sequence.}\label{fig:fig3} 
\end{figure}

The entropy distribution as a function of the state vector for all the
above three ciphers along with that from the new method is shown in
figure \ref{fig:fig3}. It is seen that the proposed method produces the
ideal distribution, a linear increase in entropy count at all bit levels in the
finite space. 

\begin{figure}
\centering {\resizebox*{0.5\textwidth}{0.3\textheight}
{\includegraphics{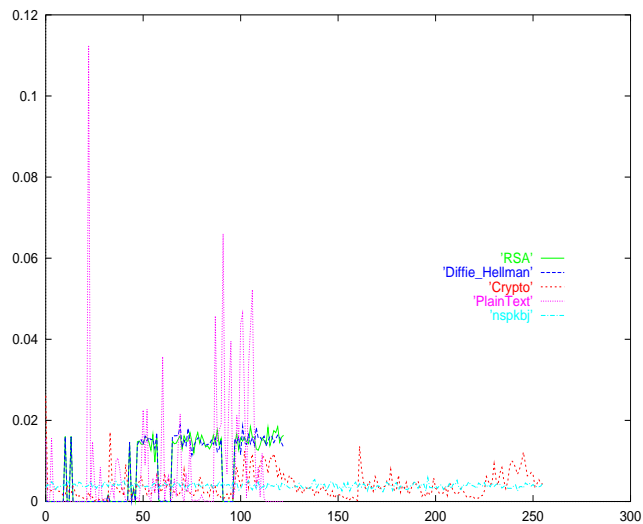}}}\caption{For an ideal cipher, the probability distribution density of the cipher
text should be independent of the frequency distribution of the plain text data.
The diagram shows the distribution of the plain text information along with
the output of the three ciphers mentioned in the text and also the one proposed.
The proposed cipher has almost ideal distribution (equal level) at all bit levels.}\label{fig:fig4}
\end{figure}

Resistance of a cipher to differential and linear cryptanalysis may be
quantitatively measured to some extent as its ability to hide any
tangible information regarding the population of the states in the plain
text. This is to say that the probability distribution density of the cipher
text should be independent of the frequency distribution of the plain text
data. In an ideal case, the probability of occurrence at all possible levels
of the iteration should be the same. This would result in a straight line
parallel to the axis of possible states of the system. Figure
\ref{fig:fig4} shows such a graph of the probability distribution density of the cipher texts
produced by the popular ciphers mentioned above along with that by the new method.
It may be noted that the new method produces a graph which is almost a straight
line as required by the ideal case. \\

\section{Conclusion}

A new algorithm for secure communication using mixing of two or more
chaotic sequences is discussed. The mixing of two near by trajectories
of the logistic equation in the fully developed chaotic scenario is shown
to generate pseudo random numbers of high complexity. The system
appears to be resistant to brute force, known plain text and differential
cryptanalysis. Since the computation is limited to finite space and
mostly XOR operations, a successful attack based on Differential Power
Analysis and Timing Attacks are not realistic on the system. Thus even
for small values of the secret key size, the cipher appears to be secure
to the popular methods of cryptanalysis. The paper also discuss some
measures to identify significant features like cycle length and
probability distribution density and compares the new cipher with some
popular ciphers.


\begin{thebibliography}{99}
\bibitem{Bras88} G. Brassard, \textit{Modern Cryptology - A Tutorial},
LNCS 325, Springer-Verlag, Berlin, 1988.  
\bibitem{Diffie79} W. Diffie, M. E. Hellman, ``Privacy and
Authentication: An introduction to Cryptography," in \textit{ Proc. IEEE}, 
vol. 67, no.3, pp. 397 - 427, 1979.
\bibitem{Berge} P. Berge, Y. Pomeav and C. Vidal, \textit{Order within
Chaos}, Hermann, Paris, France, 1986. 
\bibitem{Gleick} J. Gleick,\textit{Chaos: Making a New Science}, Viking
Penguin, New York, 1987. 
\bibitem{Nicolis} G. Nicolis and I. Prigogine, \textit{Exploring
Complexity}, W. H. Freeman and Company, New York, 1989. 
\bibitem{RitterT} T. Ritter, ``The Efficient Generation of Cryptographic
Confusion Sequences," \textit{Cryptologia}, vol. 15, no.2, pp. 81 - 139, 1991. 
\bibitem{Pecora} M. Louis Pecora and L. Thomas Carroll,
``Synchronization in Chaotic systems," \textit{Physical Review Letters},
vol. 64, no. 8, pp. 821 - 824, 1990. 
\bibitem{Hong} H. Li and J. Chern, ``Coding the chaos in a semiconductor
diode for information transmission," \textit{Physics Letters A}, vol. 206, 
pp. 217 - 221, 1995.
\bibitem{Young} Y. H. Yu, K. Kwak and T. K. Lim,
``Secure communication using small time continuous feedback,"
\textit{Physics Letters A}, vol. 197,  pp. 311 - 315, 1995. 
\bibitem{RitterT90} T. Ritter, ``Substitution Cipher with
Pseudo-Random Shuffling: The Dynamic Substitution Combiner,"
\textit{Cryptologia}, vol. 14, no. 4, pp. 289 - 303, 1990. 
\bibitem{MacLaren} M. MacLaren and G. Marsaglia, ``Uniform Random
Number Generator," \textit{Journal of Association for Computing
Machinery}, vol. 12 no. 1, pp. 83 - 89, 1965. 
\bibitem{Wolfram} S. Wolfram, ``Cryptography with Cellular Automata
(extended abstract)," in: \textit{Advances in Cryptography, CRYPTO'85
Proceedings}, Springer-Verlag,  pp. 429 - 432, 1986. 
\bibitem{Pless} V. Pless , ``Encryption Schemes for Computer
Confidentiality," \textit{IEEE Transactions on Computers},
vol. C-26, no. 11, pp. 1133 - 1136, 1977. 
\bibitem{Beck} C. Beck, \textit{Signal Processing VI: Theories and applications}, Elsevier,
Amsterdam, 1992. 
\bibitem{Cernak} J. Cernak, ``Digital generators of chaos," \textit{Physics
letters A}, vol. 214,  pp. 151 - 160, 1996.  
\bibitem{Matthews} R. Matthews, ``On the Derivation of a Chaotic
Encryption Algorithm," \textit{Cryptologia}, vol. 13, pp. 29 - 42, 1989. 
\bibitem{Wheeler} D. Wheeler, ``Problems with Chaotic Cryptosystems,"
\textit{Cryptologia}, vol. 13, pp. 243 - 250, 1989. 
\bibitem{Mitchell} D. Mitchell, ``Nonlinear Key Generators,"
\textit{Cryptologia}, vol. 14, pp. 350-354, 1990.
\bibitem{Rietman}E. Rietman, \textit{Exploring the Geometry of
Nature}, Windcrest Books, Blue Ridge Summit, P.A, 1989.
\bibitem{Siegenthaler} T. Siegenthaler, ``Decrypting a Class of Stream
Ciphers Using Ciphertext Only," \textit{IEEE Transactions on
Computers}, vol. C-34, pp.  81 - 85, (1985).
\bibitem{Knuth81} D. Knuth, \textit{The art of Computer Programming,
Vol.2, Seminumerical Algorithms}, 2nd ed., Addison-Wesley, Reading,
Massachusetts, 1981. 
\bibitem{Ritter_COMBCORR}T. Ritter, ``The story of Combination
Correlation: A Literature Survey," Research comments from Ciphers by
Ritter, (1996), URL: http://www.io.com/\( \sim  \) ritter/ under
Literature Surveys and Reviews. 
\bibitem{Fischer} E. Fischer, ``A theoretical Measure of Cryptographic
Performance," \textit{Cryptologia}, vol. 5, no.1, 1981. 
\bibitem{Biham90} E. Biham and A. Shamir,
``Differential Cryptanalysis of DES-like Cryptosystems," in:
\textit{Advances in Cryptology: Crypto'90}, Springer-Verlag, 1990.
\bibitem{Huffman} D. Huffman, ``A method for the construction of
Minimum-Redundency Codes," in: \textit{Proceedings of the IRE}, vol.
40, pp. 1098 - 1101, 1952. 
\bibitem{Heller} S. Heller, \textit{Efficient C/C++ Programming:
Smaller, Faster, Better}, 2nd Edition, chapters 1-4, Academic Press
Professional, 1991. 
\end{thebibliography}
\end{document}